\documentclass[12pt]{article}
\usepackage[dvips,bookmarks]{hyperref}
\hypersetup{colorlinks,bookmarksopen,bookmarksnumbered,citecolor=blue,
	pdfstartview=FitH}
\def\hhref#1{\href{http://arxiv.org/abs/hep-th/#1}{hep-th/#1}}
\def\mhref#1{\href{mailto:#1}{#1}}
\begin{document}
\newcommand{\p}{\partial}
\newcommand{\hp}{\hat{\p}}
\newcommand{\ov}{\overline}
\newcommand{\da}{^{\dagger}}
\newcommand{\w}{\wedge}
\newcommand{\al}{\alpha}
\newcommand{\bb}{\beta}
\newcommand{\ga}{\gamma}
\newcommand{\te}{\theta}
\newcommand{\de}{\delta}
\newcommand{\et}{\eta}
\newcommand{\ze}{\zeta}
\newcommand{\s}{\sigma}
\newcommand{\e}{\epsilon}
\newcommand{\om}{\omega}
\newcommand{\Om}{\Omega}
\newcommand{\la}{\lambda}
\newcommand{\La}{\Lambda}
\newcommand{\n}{\nabla}
\newcommand{\hn}{\hat{\nabla}}
\newcommand{\hph}{\hat{\phi}}
\newcommand{\ad}{\dot{a}}
\newcommand{\bd}{\dot{b}}
\newcommand{\gd}{\dot{c}}
\newcommand{\dd}{\dot{\delta}}
\newcommand{\ed}{\dot{\eta}}
\newcommand{\zd}{\dot{\zeta}}
\newcommand{\md}{\dot{m}}
\newcommand{\nd}{\dot{n}}
\newcommand{\tb}{\overline{\theta}}
\newcommand{\ti}{\widetilde}

\newcommand{\2}{\textstyle{1\over 2}}
\newcommand{\3}{\textstyle{1\over 3}}
\newcommand{\4}{\textstyle{1\over 4}}
\newcommand{\8}{\textstyle{1\over 8}}
\newcommand{\6}{\textstyle{1\over 16}}
\newcommand{\ra}{\rightarrow}
\newcommand{\Ra}{\Rightarrow}
\newcommand{\im}{\Longleftrightarrow}
\newcommand{\hs}{\hspace{5mm}}
\newcommand{\x}{\star}
\newcommand{\Delt}{\p^{\star}}

\thispagestyle{empty}
{\bf 12th March, 2002} \hspace{\fill}
{\bf YITP-SB-02-09}

\vspace{1cm}
\begin{center}{\Large{\bf RADIAL DIMENSIONAL REDUCTION:\\

\vspace{3mm}
 (ANTI) DE SITTER THEORIES FROM FLAT}}\\
\vspace{1cm}
{\large{\bf T. Biswas\footnote{\mhref{tirtho@insti.physics.sunysb.edu}} and W. Siegel\footnote{\mhref{siegel@insti.physics.sunysb.edu}}}}\\
\vspace{5mm}
{\small C.N. Yang Institute for Theoretical Physics\\
Department of Physics and Astronomy\\
State University of New York at Stony Brook\\
Stony Brook, New York 11794-3840}
\end{center}

\begin{abstract}
We propose a new form of dimensional reduction that constrains dilatation instead of a component of momentum. It corresponds to replacing toroidal compactification in a Cartesian coordinate with that in the logarithm of the radius. Massive theories in de Sitter or anti de Sitter space are thus produced from massless (scale invariant) theories in one higher space or time dimension. As an example, we derive free massive actions for arbitrary representations of the (anti) de Sitter group in arbitrary dimensions. (Previous general results were restricted to symmetric tensors.) We also discuss generalizations to interacting theories. 
\end {abstract}

\newpage
\setcounter{page}{1}
\section{INTRODUCTION} 
Consistent generally covariant actions are known to exist only for fields with spin less than or equal to two. However, we know that string theory contains an infinite tower of states with all sorts of higher spin representations. Thus for example, to establish a connection between the superstring theory and the low energy effective field theories (for these higher spin particles) one needs to make progress in understanding  higher spin gauge field theories. A natural first step towards this would be to construct gauge invariant actions for  fields (of arbitrary representations) in Anti-de Sitter (AdS) background. AdS is compatible with supersymmetry (see for eg. \cite{adssusy}) and interacting massless fields can consistently propagate \cite{adsvacuum} in it. Studying  fields (field equations and actions) of arbitrary representations in anti-de Sitter or de Sitter (dS) background is also important in the context of AdS/CFT  correspondence \cite{adscft}. However, this task has eluded physicists for a long time although progress have been made in specific dimensions and/or for specific   representations\footnote{Bosonic actions for symmetric representations in dS/AdS background have been obtained, first in four dimensions \cite{ads4bosact} followed by the massless case in arbitrary dimensions \cite{vasbosact}  and recently the massive case in arbitrary dimensions \cite{adsbosact}. A recent paper \cite{ads5bosact} has also obtained the lightcone gauge action for arbitrary massless bosonic representations in five dimensions. The story with fermions is bleaker and till now actions have only been obtained for symmetric massless representations \cite{ads5feract}.}.

In this paper we present a simple and elegant solution to this problem in its full generality and then work out some examples. First we obtain the field equations for the field strengths (of gauge fields) of arbitrary representations in (Anti-)de Sitter space-time of arbitrary dimension and then  construct their gauge invariant action, both by what we call ``radial dimensional reduction'' of one higher dimensional flat space-time. Further, we believe that this prescription relates theories in flat space-time with that of dS or AdS space-time in a much more general sense. The earliest effort in trying to relate theories on dS/AdS space-time with that of flat space-time was by Dirac \cite{dirac} where he discussed  field equations for scalars, vectors and spin half fermions. 

Results for massless multiplets in flat space-time \cite{flataction} were derived using  group (conformal) theoretic techniques and first quantized BRST methods. Our efforts to directly apply these procedures to dS/AdS space-time failed and we realised that perhaps we have to invoke extra dimensions to make it work. We therefore set out to try and mimic the construction of massive flat space-time multiplets\footnote{By the term multiplet we  mean a collection of fields which together describe a representation of the relevant symmetry group which in our case is the D dimensional dS/AdS.} which were obtained by dimensional reduction of the massless flat space-time multiplets in one higher dimension \cite{flatmassive}. At first sight the generalization to the dS/AdS case  may appear impossible because we did not have the massless dS/AdS multiplets to begin with. However we know that dS/AdS in say D dimensions can adequately be described as a hypersurface (``sphere'') in D+1 dimensional flat space-time with appropriate signature. Thus we can start with a D+1 dimensional flat space-time but instead of dimensionally reducing a flat direction (which would give us the massive flat space-time multiplets) we can  reduce the radial coordinate which should then give us dS/AdS. This geometric picture was the basic motivation behind our construction and it works! 

A more detailed analysis, however, reveals that in order that this procedure work for both field equations and action we need  another crucial ingredient, that of scale invariance,  and this is where geometry gives way to algebra. For the reduction of the field equations to work we need a generator which commutes not only with the isometry group of the dS/AdS hypersurface but also with the field equations. Indeed we have such a generator, the dilatation, which is not the same as the radial momentum as one would naively expect but closely related to it. A simple way to reduce actions  is to compactify the extra coordinate. If the original action is scale invariant then the various ``harmonic modes'' (corresponding to the  eigenfunctions of the dilatation) decouple from one another to give us separate D dimensional AdS/dS actions corresponding to each mode.  Fortunately, the massless flat D+1 dimensional gauge invariant action is also scale invariant. It is clear then that instead of choosing eigenspaces of the radial momentum operator ($\sim \p/\p r$) to reduce the Hilbert space (to D dimensions) we have to choose eigenspaces of the dilatation ($\sim r\p/\p r$). Dilatation now plays the role that the momentum generator in the extra flat direction plays in the usual compactification scheme. 
\begin{equation}
 \Delta = im \ \hbox{\hspace{5mm} instead of \hspace{5mm}} \ P_D = m 
\end{equation}
Thus in the comparison between the usual reduction scheme and our scheme, the appropriate analogue of $x^{D}$, the extra flat coordinate is $u=ln(r)$ and not $r$. Accordingly, the appropriate ``radial compactification'' in analogy with the usual ``toroidal compactification'' will be given by
$$ e^{2\pi\Delta/m} = 1 \ \hbox{\hspace{5mm} instead of \hspace{5mm} } \ e^{2\pi i P_D/m} = 1 $$ 

Through radial dimensional reduction we obtain both the field equations and action for the various field representations. We observe that the fields which start out as representations of SO(D,1) or SO(D-1,2) contains a set of several irreducible representations (irreps.) of SO(D-1,1) or SO(D-1,1) respectively for dS/AdS. Another way of saying this is that massive dS/AdS   multiplets contain several irreps.\ of flat space-time, which corroborates our previous understanding. Of course for some special mass values  these multiplets may decouple into smaller multiplets \cite{deser,adsmass,adsbosact} as we shall see in the examples.   
  
In section 1, we explain the procedure of radial dimensional reduction in more detail. In section 2, we obtain the field equations for the field strengths of gauge fields in D dimensional dS/AdS background by dimensional reduction of the D+1 dimensional flat space-time field equations. We also derive the lightcone form of the AdS generators by solving the field equations and imposing the light-cone constraints,  and check that it agrees with the already known results \cite{adslc}. In section 3, we obtain the gauge invariant action (for both fermions and bosons) in dS/AdS background, and  in section 4 we work out some simple examples for which results are already known. In section 5 we briefly discuss possible generalizations of radial dimensional reduction to interacting theories. Finally we conclude by commenting on possible future research.  

\section{RADIAL DIMENSIONAL REDUCTION}
{\bf The Generic Structure:} 
Our starting point will be the results that were derived  for massless multiplets in flat space-time which was first obtained for symmetric representations for both bosons \cite{flatbosact} and fermions \cite{flatferact} and later on generalised to arbitrary representations \cite{flataction} using BRST methods. A straightforward attempt to generalise the flat space-time derivation \cite{flataction} for dS/AdS fails. For example, we can indeed obtain field equations that are invariant under the dS/AdS symmetry group but their algebra does not close (i.e. the commutators of the field equations produce newer ones). After adding sufficient field equations, so that the algebra closes, we end up getting the conformal field equations, which describe only the conformal multiplets! The failure  made us  realise that perhaps we have to introduce extra dimensions to make the group theoretic methods work, just as one has to in order to obtain massive multiplets in flat space-time. Let us then investigate the general scheme of dimensional reduction. 

Suppose we want to obtain the field equations and action for fields on a D dimensional maximally isometric manifold $\Om$ which can be described as a hypersurface in a D+1 dimensional flat space-time with appropriate signature. Our first objective then is to find  field representations of the isometry group ($J$) of $\Om$. We can achieve  this by dimensional reduction if
\begin{description}
\item[I.] We have a group $\hat{J}$ isomorphic to the isometry group of $\Om$ with generators $\{\hat{J}_i\}$ acting on the D+1 dimensional flat manifold which leave the hypersurface $\Om$ invariant.
\end{description}

This condition implies that the tangent vector (as an operator), say $E$, which is normal to the hypersurface $\Om $, commutes with the generators $\{\hat{J}_i\}$. To make matters concrete  let us choose a coordinate system $\{x^m,e\}$ of the D+1 dimensional flat space such that $x^m$, $m=0 ... D-1$, labels the points in $\Om$ while $e$ is the coordinate which is constant in $\Om$. Then the eigenfunctions of $E$ given by
\begin{equation}
E\phi=\frac{\p}{\p e}\phi=\la\phi
\end{equation}
for a fixed $\la$ (lets call the eigenspace $H_{\la}$) forms a representation of $\hat{J}$.
This representation of $\hat{J}$ is still in terms of D+1 dimensional fields but one can easily expand in the eigenstates of $E$, so that we have a one to one mapping between the eigenfunctions of $E$ in $H_{\la}$ and D dimensional functions living on the hypersurface $\Om$.  Hence we now have a representation of $J$ (since $\hat{J}$ and $J$ are isomorphic) in terms of D dimensional fields on $\Om$.

Our next task is to find a set $F$ of field equations $\{F_I\}$ for field strengths of the fields. It should satisfy the following requirements:
\begin{description}
\item[a.] The algebra of $F_I$'s should close.
\item[b.] $F$ should be invariant under $J$, i.e. $[F_I,J_i]\sim F_J$
\end{description}

Now suppose we already have a set $\hat{F}$ of field equations $\{\hat{F}_I\}$ for D+1 dimensional field-representations of $\hat{J}$. We can perform a dimensional reduction of $\hat{F}$ similar to that of $\hat{J}$ if $\hat{F}$ commutes with $E$ as then $\hat{F}$  can be defined on the hypersurface $\Om$. Also, because $\hat{F}$ satisfy (a) and (b) for D+1 dimensional flat space they automatically satisfy them for $\Om$. (Remember $\hat{J}$ and $J$ has been identified.) Thus we need the following further hypotheses to make the field equations work:
\begin{description}
\item[IIA.] There should exist a set of field equations $\hat{F}=\{\hat{F}_I\}$ whose algebra closes.
\item[IIB.] $\hat{F}$ should be invariant under $\hat{J}$, i.e. $[\hat{F}_I,\hat{J}_i]\sim \hat{F}_J$
\item[IIC.] $\hat{F}$ commutes\footnote{Actually it turns out that this is a stricter requirement then necessary. Even if $E$ leaves $\hat{F}$ invariant we can find another suitable set $\hat{F}'$ which will commute with $E$ and satisfies all other requirements.} with $E$ or in other words $\hat{F}$ leaves $\Om$ invariant.
\end{description}

Next we turn our attention to the action. Again, if already we have a gauge invariant action for D+1 dimensional field representations which is also invariant under $\hat{J}$, then after dimensional reduction we should get an action which is still gauge invariant and ``$J$-invariant''. However, for this to be true  the partial integrations that we need to perform (to prove gauge invariance for example) should still work, even for eigenfunctions of $E$ which do not necessarily vanish at $r=0,\infty$. The most natural way to ensure that the total derivative terms vanish is to compactify the extra coordinate $e$:
$$e\approx e+2\pi$$
Since the original action is invariant under $E=\p/\p e$ the various harmonic modes   decouple in the action and we obtain separate actions for each mode. These modes now live in the D dimensional hypersurface $\Om$. Note for us $e$ does not have any physical significance (unlike the  extra dimensions in Kaluza-Klein reduction) and just serves the mathematical purpose of providing us with actions on $\Om$ and hence for example, the relative or overall co-efficients in front of the various ``mode actions'' is unimportant. The gauge transformations of the various modes can be easily obtained by harmonic expansion of the original gauge transformation.  Thus for the dimensional reduction of the action to work we further need\footnote{We do not need II for the actions while III is obviously irrelevant for the the field equations.}:
\begin{description}
\item[III.] D+1 dimensional flat gauge invariant action to be invariant under $E$.
\end{description}
\vspace{5mm}
{\bf D dimensional dS/AdS space-time:} With the general understanding of how the dimensional reduction works we can now specialise to the case of  reducing D+1 dimensional flat space-time with signature $(-+...++)/(-+...+-)$ to D dimensional dS/AdS. In this case the role of $e$ is played by the ``radial coordinate'' $r$, or more precisely $u\equiv ln(r)$, while $E$ is the dilatation generator $\Delta$. To make things clear let us choose a particular coordinate system:
$$
y^m=\frac{\sqrt{2}rx^m}{(1+\2\eta x^2)};\ \ y^D=\frac{r(1-\2\eta x^2)}{(1+\2\eta x^2)}$$
\begin{equation}
\hat{\eta}_{MN}=\left( \begin{array}{cc} \eta_{mn}&0\\
                                             0 &\eta
\end{array} \right)
\end{equation}
where $y^M$ are the flat D+1 dimensional coordinates with metric $\hat{\eta}_{MN}$. In the metric $\eta=+1$ corresponds to dS, while $\eta=-1$ corresponds to  AdS.
To be sure one can easily compute the metric element in this coordinate system:
\begin{equation}
ds^2=\eta dr^2 + 2r^2\frac{\eta_{mn}}{(1+\2\eta x^2)^2}dx^mdx^n 
\end{equation}
Clearly the $\Om$ hypersurface has the familiar metric of dS/AdS space-time. As perhaps now one can guess that the role of $\hat{J}$ will be played by the Lorentz generators of the D+1 dimensional flat space-time. As usual the Lorentz generators have an orbital piece and a spin piece. (In our paper we refer to the spin piece as spin operators, and by ``spin-fields'' we refer to the   fields which are representations of these  D+1 dimensional spin operators.  The spin value of the spin-fields then corresponds to the Casimir operator $S^2= S^{AB}S_{AB}$.) We know that $\hat{J}=SO(D,1)$ or $SO(D-1,2)$ for $\eta=\pm 1$, and is precisely the isometry group of dS/AdS. These obviously commute with the dilatation $\Delta$ and thus condition (I) is satisfied. To see how the Lorentz generators look after reduction one needs the inverse coordinate transformation:
\begin{equation}
r^2=\eta y^2;\  x^m=\frac{\sqrt{2}y^m}{(y^D+r)}
\end{equation}
One can then compute the partial derivatives:
$$
\frac{\p}{\p y^m}=\frac{1}{\sqrt{2}r}\left[\de_m^n(1+\2\eta x^2)-\eta x^n x_m\right]\frac{\p}{\p x^n}+\frac{\sqrt{2}\eta x_m}{(1+\2\eta x^2)}\frac{\p}{\p r} 
$$and 
\begin{equation}
\frac{\p}{\p y^D}=-\frac{x_m}{r}\frac{\p}{\p x^m}+\frac{(1-\2\eta x^2)}{(1+\2\eta x^2)}\frac{\p}{\p r}
\end{equation}
We know the orbital piece (which corresponds to the Killing vector) of the Lorentz generators in terms of $y^M$ coordinates:
\begin{equation}
L_{MN}=y_{[M}\frac{\p}{\p y^{N]}}
\end{equation}
Then using (5),(6) and (7) we find
\begin{equation}
L_{mn}=x_{[m}\frac{\p}{\p x^{n]}} ;\hspace{5mm} L_{Dm}=\frac{\eta}{\sqrt{2}}\left[\frac{\p}{\p x^m}-\frac{1}{2}\eta\left(x^2\frac{\p}{\p x^m}-2x_mx\cdot\p\right)\right]
\end{equation}
One immediately recognises that the orbital pieces of the generators correspond to the Killing vectors for dS/AdS space-time (in the given coordinate system\footnote{In this coordinate system the Killing vectors are given by the orbital piece of $J_{mn}$, the D dimensional Lorentz rotations, and $(P_m-\2\eta K_m)$, where $P_m$ is the momentum generator and $K_m$ the conformal boost.}) as we expected. The Killing symmetry generators associated with the Killing vectors are then uniquely given by 
\begin{equation}
J_{mn}=x_{[m}\frac{\p}{\p x^{n]}}+S_{mn} ;\hspace{5mm} J_{rm}=\frac{\eta}{\sqrt{2}}[\frac{\p}{\p x^m}-\2\eta(x^2\frac{\p}{\p x^m}-2x_mx.\p-2 x^nS_{mn})]
\end{equation}
Clearly these are the dS/AdS symmetry generators and they correspond to the full D+1 dimensional Lorentz generators\footnote{Depending on the particular coordinate system, we have to perform different local Lorentz transformations to obtain the Lorentz generators in the original frame $J_{MN}$ in terms of the ones in the new frame ($J_{mn},J_{rm}$).}.

Thus we now have a representation of the dS/AdS  isometry generators given by (9). Note that the generators are independent of the radial coordinate, which is as it should be.
\vspace{5mm}
\\
{\bf The covariant derivative:} Now that we know that  D dimensional dS/AdS can be described as a ``sphere'' in D+1 dimensional flat space-time, we can choose an arbitrary coordinate system on the sphere (to keep things covariant on $\Om$) and proceed to implement the radial dimensional reduction of field equations and actions. Since we are now working in a more general coordinate system, we will  need the covariant derivatives instead of the partial derivatives (6). In this subsection we compute them.  

In this more general coordinate system the metric element looks like:
\begin{equation}
ds^2=\eta dr^2 + r^2g_{mn}dx^mdx^n
\end{equation}
where $g_{mn}$ is now the dS/AdS metric in any arbitrary coordinate system. It is simpler to work in the vielbein formalism. We choose our vielbein and its inverse as follows:
\begin{equation}
E_M{}^{A}=\left( \begin{array}{cc}
re_m{}^{a} & 0\\
0 &1
\end{array} \right);
\hspace{5mm}
E_A{}^{M}=\left( \begin{array}{cc}
\frac{1}{r}e_a{}^{m} & 0\\
0 &1
\end{array} \right)
\end{equation}
where $e_m{}^{a}$ is the vielbein for dS/AdS space time and  the ``flat metric'' is 
\begin{equation}
\eta_{AB}=\left( \begin{array}{cc}
\eta_{ab} & 0\\
0 &\eta
\end{array} \right)
\end{equation}

The covariant derivative in general is given by
\begin{equation}
\nabla_A=E_A+\2\om_A{}^{BC}M_{CB}
\end{equation}
where $E_A=E_A{}^{M}\p_M$, $\om_A{}^{BC}$ are the spin connections and $M_{CB}$ are the second quantized spin generators \cite{flataction}. However instead of using (13), for the simple vielbein  matrix (11) it is  easier to compute the covariant derivatives using Weyl transformations which is given by
\begin{equation}
E_A{}^M(x^M)=W(x^M)E'_A{}^M(x^M);\hspace{5mm} \n_A=W\n'_A+(\n'^BW)M_{AB}
\end{equation}
where $W(x^M)$ is the Weyl gauge parameter. 
Now consider our case:
$$
E_A{}^{M}(x^m,r)=\left( \begin{array}{cc}
\frac{1}{r}e_a{}^{m}(x^m) & 0\\
0 &1
\end{array} \right)
= \frac{1}{r}\left( \begin{array}{cc}
e_a{}^{m}(x^m) & 0\\
0 &r
\end{array} \right)\equiv W(r)E'_A{}^M(x^m,r)
$$
Using (14) we then obtain
\begin{equation}
\n_A=\frac{1}{r}\n'_A+\left(\n'^B\frac{1}{r}\right)M_{AB}
\end{equation}
Now the vielbein $E'_A{}^M$ is a direct product of the vielbeins for the D dimensional dS/AdS and the one-dimensional ``$r$'' space. Thus the covariant derivatives $\n'_a$ are just the covariant derivatives for dS/AdS while $\n'_r=r\p/\p r$. 
Then
\begin{equation}
\n'^B\frac{1}{r}=-\de_r{}^B\eta\frac{1}{r}
\end{equation}
From now on ``hatted'' quantities will denote D+1 dimensional objects, while ``unhatted'' will denote D dimensional objects. Thus we have
\begin{equation}
\hat{\n}_{a}=\frac{1}{r}(\n_a+\eta M_{ra})
\end{equation}
\begin{equation}
\hat{\n}_{r}=\frac{\p}{\p r}
\end{equation}

It is also convenient to work with the ``Weyl transformed'' fields, $\tilde{\phi}(r,x^m)$:
\begin{equation}
\tilde{\phi}(r,x^m)\equiv (W(r))^{-d}\hph(y^M)=r^{d}\hph(y^M)
\end{equation}
where $d$ is the scale dimension for the fields in D+1 dimensions. Then the action of the covariant derivatives on the original fields can be written as
\begin{equation}
\hat{\n}_{a}\hph(y^M)=r^{-d}\frac{1}{r}(\n_a+\eta M_{ra})\tilde{\phi}(r,x^m)
\end{equation}
\begin{equation}
\hat{\n}_{r}\hph(y^M)= r^{-d}\frac{1}{r}(-d+\Delt)\tilde{\phi}(r,x^m)
\end{equation}
where we define
\begin{equation}
\Delt\equiv \frac{\p}{\p u}=r\frac{\p}{\p r}
\end{equation}

\section{FIELD EQUATIONS AND LIGHT-CONE\\ GENERATORS}
{\bf Field Equations for Field strength:} We already have field equations in D+1 flat space-time in flat coordinates \cite{flataction}:
\begin{equation}
\hp^2=0;\ \ \om \hp_A+S_A^{\ B}\hp_B=0
\end{equation}
where 
\begin{equation}
\om=d-\frac{D-1}{2}
\end{equation}
 First we covariantize (23) to obtain the equations for flat space-time in a general coordinate system:
\begin{equation}
\hat{F}\equiv\hn^2=0;\ \ \hat{F}_M\equiv\om \hn_A+S_A^{\ B}\hn_B=0
\end{equation}
These equations are not only invariant under $\hat{J}$ satisfying (IIA) with a closed algebra (IIB) but also invariant under the scale transformation $\Delta$. However, for the dimensional reduction to work we need them to commute with $\Delta$ (IIC). By multiplying the field equations by appropriate scale factors of ``$r$'' this can be achieved and one can easily verify that this scaling does not disturb (IIA) and (IIB). It turns out that the correct field equations\footnote{One can derive these field equations more directly: Lorentz invariance just implies that the field equations have the right tensor structure. The simplest choice being $\hat{F}=f(y)\hp^2$ and $ \hat{F}_M= f_1(y)(\om(y) \hp_A+S_A^{\ B}\hp_B)$. Then the fact that the algebra should close and that for the dimension reduction to work the field equations must commute with $\Delta$ gives the precise values for the arbitrary functions $f(y),f_1(y)$ and $w(y)$.} are given by
\begin{equation}
\hat{F}=r^2\hn^2;\ \ \hat{F}_M=r(\om \hn_A+S_A^{\ B}\hn_B)
\end{equation}
Now to dimensionally reduce these field equations we have to find out  how they act on the eigenspace $H_{m}$ where the eigenfunctions satisfy the eigenvalue equation
\begin{equation}
\Delt \tilde{\phi}(u,x)=im\tilde{\phi}(u,x) \ \mbox{ or }\  \Delta \hat{\phi}(r,x)=im\hat{\phi}(r,x)
\end{equation}
 and hence look like
\begin{equation}
\tilde{\phi}(u,x)=e^{imu}\phi(x)\ \mbox{ or }\ \hat{\phi}(r,x)=r^{-d+im}\phi(x)\equiv r^{\la}\phi(x)
\end{equation}
One can now see how the ``field equation operators'' look in the D dimensional space-time $\Om$ from their action on $H_{m}$:
\begin{equation}
\hat{F}\hph(y)=\hat{F}r^{-d+im}\phi(x)\equiv r^{-d+im}F\phi(x)
\end{equation}
where $F$ thus obtained is  the corresponding ``field equation operator'' in the D dimensional $\Om$. Similarly we can get $F_M$ from $\hat{F}_M$. We now proceed to obtain them explicitly.
$$\hat{F}=r^2\hn^2=r^2(\hn^a\hn_a+\hn^r\hn_r)$$
Let us evaluate $\hn^a\hn_a$ first.
$$\hn^a\hn_a=\frac{1}{r}(\n^a+\eta M_r{}^{a})\hn_a$$
(substituting $\hn^a$ from (17))
$$ =\frac{1}{r}(\n^a\hn_a+\eta D \hn_r+\eta \hn_aM_r{}^{a})$$
(since $M_r{}^{a}\hn_a=[M_r{}^{a},\hn_a]+\hn_aM_r{}^{a}=D\hn_r+\hn_aM_r{}^{a}$)
$$=\frac{1}{r^2}(\n^2+2\eta \n^aM_{ra}+\eta D\la + M_{ra}M_r{}^{a})$$
Next we evaluate $\hn^r\hn_r$.
$$\hn^r\hn_r=\eta\frac{\p}{\p r}\frac{\p}{\p r}=\frac{\eta}{r^2}\la(\la-1)$$
Thus we finally have
$$F=\n^2-2\eta \n^aS_{ra}+\eta \la(D+\la-1) + S_{ra}S_r{}^{a}$$
where we have used $M\hat{\phi}=-S\hat{\phi}$, the usual relation between the first and second quantized operators. Using the definition of $\la$ we get
\begin{equation}
F=\n^2-2\eta \n^aS_{ra} + S_{ra}S_r{}^{a}+\eta(2d\om-m^2-d^2-2im\om)
\end{equation}
Next let us look at $\hat{F}_a$.
$$\hat{F}_a=r(\om\hn_a+S_a{}^{B}\hn_B)$$ Now
$$\om\hn_a+S_a{}^{B}\hn_B=\om\hn_a+S_a{}^{b}\hn_b+S_a{}^{r}\hn_r $$ 
$$=\frac{1}{r}(\om\n_a+S_a{}^{b}\n_b-\eta\om S_{ra}-\eta S_a{}^{b}S_{rb} +\la S_{ar})$$
Then
\begin{equation}
F_a=\om\n_a+S_a{}^{b}\n_b-\eta(\om -d+im)S_{ra}-\eta S_a{}^{b}S_{rb}
\end{equation}
We are still left with $\hat{F}_r$.
$$\hat{F}_r=r(\om \hn_r+S_r{}^{A}\hn_A)$$ 
Now
$$\om \hn_r+S_r{}^{A}\hn_A=\om \hn_r+S_r{}^{a}\hn_a=\frac{1}{r}(\om\la+S_r{}^{a}\n_a-\eta S_r{}^{a}S_{ra})$$
Thus
\begin{equation}
F_r=S_r{}^{a}\n_a-\eta S_r{}^{a}S_{ra}+\om(-d+im)
\end{equation}
The complete set of field equations $\{F,F_a,F_r\}$ are then given by (30),(31) and (32).
We should mention that although field equations for most massless gauge fields were known for both bosons \cite{adsboseqn} and fermions \cite{adsfereqn}, this simple derivation is completely new and we believe that the field equations for the field strength is new, too.
\vspace{5mm}
\\
{\bf dS vs. AdS and Unitarity:} We obtained  the field equations (30)-(32) by dimensional reduction of  an extra space-like or time-like coordinate corresponding to dS or AdS respectively. It is well known that even in ordinary dimensional reduction of a time-like coordinate we run into tachyons, which render the theory nonunitary. As expected therefore, in the case of AdS when we are reducing the time-like radial coordinate, we encounter tachyons. This is clearer when one tries to obtain the AdS action by the dimensional reduction prescription. We can however circumvent this problem by performing a Wick rotation in the ``$u$'' coordinate,  which implies for example that instead of using $e^{imu}$ modes we should use $e^{mu}$ (see section 3 for details). For field equations, this essentially means that for AdS we have to replace $im$ by $m$, but the effect of Wick rotation is more subtle when we deal with actions  as we will explain in more detail in the next section. Thus the field equations for AdS look like
\begin{equation}
F=\n^2+2 \n^aS_{ra} + S_{ra}S_r{}^{a}-(2d\om+m^2-d^2-2m\om)
\end{equation}
\begin{equation}
F_a=\om\n_a+S_a{}^{b}\n_b+(\om -d+m)S_{ra}+ S_a{}^{b}S_{rb}
\end{equation}
\begin{equation}
F_r=S_r{}^{a}\n_a+ S_r{}^{a}S_{ra}+\om(-d+m)
\end{equation}

For dS the field equations are given by (33)-(35) and all we have to do is to substitute $\eta=1$.
\vspace{5mm}
\\ 
{\bf Light-cone Generators for Bosons in AdS:} light-cone form of generators is useful for various reasons. Firstly, knowing its form one can apply BRST techniques  to obtain the action for the spin gauge fields. This was done successfully for example to derive the flat space-time gauge invariant action \cite{flataction}. Secondly, even though we have obtained the field equations it is only for the gauge invariant field strengths. However we are usually  interested in the physical (gauge) fields. In the light-cone formalism one eliminates the extra degrees of freedom and deals only with the physical fields. For example, the light-cone form of the symmetry generators gives us a representation in terms of only the physical fields. Such an approach might prove especially useful if a consistent covariant formalism is absent (a notable example being the superstring). Finally and most importantly for us, the light-cone form of generators for arbitrary bosonic spin representations has already been worked out for AdS \cite{adslc} purely from algebraic considerations. We therefore work out the bosonic case for AdS which provides us with a non-trivial check of our results and approach in general. 

To describe bosons we can choose $\om=0$ which implies
\begin{equation}
d=\frac{D-1}{2}
\end{equation}
and simplifies the field equations. Thus in AdS we get after minor manipulations and recombinations
\begin{equation}
F'=\n^2-S_r^{\ a}S_{ra}+(d^2-m^2)
\end{equation}
where $F'=F-2F_r$
\begin{equation}
F_a=S_a^{\ b}\n_b+mS_{ra}+\2\{S_a^{\ b},S_{rb}\}
\end{equation}
and
\begin{equation}
F_r=S_r^{\ a}\n_a+S_r^{\ a}S_{ra}
\end{equation}

There are various ways to obtain the light-cone generators. Our approach will be to first impose constraints on the Hilbert space, which removes extra degrees of freedom (gauge degrees).  Then we solve the field equations (at the level of operators) which then effectively reduces the big Hilbert space (of field strengths) to  the Hilbert space of physical states involving only gauge (physical) fields. To apply this procedure however we have to work in a specific coordinate system. We choose the coordinate system where the AdS vielbein and its inverse looks like
\begin{equation}
e_m^{\ a}=\frac{1}{z}\de_m^{\ a}; \ \ 
e_a^{\ m}=z\de_a^{\ m}
\end{equation}
where $m=0...D-1$ and the coordinates are given by $(x^{\md},z)$, $\md=0...D-2$ with $\eta_{zz}=1$.
In this frame one can compute the field equations (see appendix for details) which after some re-organization and appropriate scalings (by factors of $z$) are given by
\begin{equation}
\tilde{F}=z^2(\p^2+\p_z\p_z)-A
\end{equation}
\begin{equation}
\tilde{F_{\ad}}=zS_{\ad}^{\ b}\p_b+\2[S_{\ad z},A]
\end{equation}
\begin{equation}
\tilde{F_z}=zS_z^{\ \bd}\p_{\bd}+B
\end{equation}
where $A$ and $B$ are operators given entirely in terms of the spin operators $S_{AB}$:
\begin{equation}
A=p^2+\de^2-2m\de+m^2-\4;\ \ B=\2p^2+k\cdot p+\de(-m+\frac{D-1}{2})
\end{equation}
where we have introduced the operators $(p_{\ad},k_{\ad},\de)$ as linear combinations of the  spin operators\footnote{We introduced these new operators  to facilitate comparison with the already existing results at the risk of making things a little abrupt perhaps.}:
\begin{equation}
p_{\ad}=S_{z\ad}+S_{r\ad};\ \ k_{\ad}=\2(S_{z\ad}-S_{r\ad});\ \ \de=S_{zr}
\end{equation}

We now go into the light-cone frame  where the flat indices further break up into $\ad=i,+,-$  the usual way; $i$ now represent the transverse directions, $i=1...D-3$. This essentially involves the coordinate transformation 
\begin{equation}
x^{\pm}=\textstyle{1\over \sqrt{2}}(x^{D-2}\pm x^0)
\end{equation}
and the metric now is given by
\begin{equation}
\eta_{ij}=1;\ \ \eta_{+-}=\eta_{-+}=1
\end{equation}
To obtain the physical space we impose the following operator constraints (gauge conditions)
\begin{equation}
x^+=S_A^{\ +}=0
\end{equation}
This only changes the operators $A$ and $B$ so that  now they contain only  the transverse spin operators, effectively reducing the dimension by 2.
$$A\ra p^2+\de^2-2m\de+m^2-\4$$
\begin{equation}
B\ra \2p^2+k\cdot p+\de(-m+\frac{D-3}{2});\hspace{5mm} [S_{iz},A]\ra[S_{iz},A]
\end{equation}
where the inner products are now  with respect to only the transverse directions. We can now solve the field equations (41)-(43) which yields
\begin{equation}
P^-=\frac{1}{2P^+}(-P^2-P_z^2+\frac{A}{z^2})
\end{equation}
\begin{equation}
S^{i-}=-\frac{1}{P^+}(\frac{1}{2z}[S^{iz},A]+S^{ij}P_j+S^{iz}P_z)
\end{equation}
\begin{equation}
S^{z-}=-\frac{1}{P^+}(\frac{1}{z}B+S^{zi}P_i)
\end{equation}
Substituting these ``constraints'' in the symmetry generators one can now obtain their light-cone representation. Here we present the Lorentz generators:
$$
J^{ij}=x^{[i}P^{j]}+S^{ij}
$$
$$J^{+i}=-x^iP^+$$
$$J^{+-}=-x^-P^+$$
\begin{equation}
J^{-i}=x^-P^i-x^iP^-+\frac{1}{2zP^+}[S^{iz},A]+\frac{1}{P^+}(S^{ij}P_j+S^{iz}P_z)
\end{equation}
   
These results coincide exactly with the known results \cite{adslc} and one can verify this for  all other conformal generators in a similar way. 

\section{ACTIONS IN dS/AdS SPACE-TIME}
\noindent
{\bf Compactification and Wick Rotation:}  
We have so far successfully dimensionally reduced the symmetry generators and field equations from D+1 dimensional flat space-time to D dimensional dS/AdS space-time. We next venture to do the same for the actions. It should perhaps be mentioned that we could have also followed the prescription of obtaining the action from the light-cone form of the Lorentz generators in dS/AdS space-time just as it was done for flat space-time \cite{flataction} using BRST techniques. However, this would firstly be more cumbersome and secondly give us an answer only in a special (light-cone) coordinate system, which we have to then covariantize. It is simpler and more elegant to obtain the dS/AdS actions by performing the radial dimensional reduction of the action directly, which will also demonstrate the general applicability  of the method.  

As we mentioned in the earlier sections, we will obtain the dS/AdS gauge invariant actions by compactifying the ``$u$'' coordinate. We thus start with the known gauge invariant D+1 dimensional flat space-time action, which is typically of the form
\begin{equation}
\hat{S}=\int dy\ \hat{\phi}^{\da}(\hat{O}\hat{\phi})
\end{equation}
where $\hat{O}$ is an operator (quadratic in derivatives for bosons, and linear for fermions). This action is invariant under the (infinitesimal) Lorentz transformations 
\begin{equation}
\hph'(y)=(1+\epsilon\hat{J})\hph(y)
\end{equation}
which after compactification translates to invariance under the isometry group of dS/AdS:
\begin{equation}
\phi'(x)=(1+\epsilon J)\phi(x)
\end{equation}
The flat action (54) will also in general possess some gauge invariance which is preserved after compactification\footnote{One has to be careful with respect to total derivative terms which need to vanish to preserve gauge invariance, especially when the derivative is with respect to the extra coordinate. However, since now we have a compact radial direction, no funny boundary terms can arise.}, the mode gauge transformation laws following from the harmonic expansion of the original gauge field. 

It is convenient to view the compactification scheme in two steps. First we perform a Weyl transformation (19). Then the action looks like
$$S=\int dx\ e\int dr\ r^{(D+2\la-\s)} \tilde{\phi}^{\da}(O\tilde{\phi})$$
where $\hat{O}r^{-d}\equiv r^{-d-\s}O$. For a bosonic action, as we shall see later $\s=2$ while for the fermions $\s=1$. Further we know that for bosons $\om=0\Ra d=\2(D-1)$ while for fermions $\om=\2 \Ra d=\2 D$ and thus in both cases we have
\begin{equation}
S=\int dx\ e{\cal L};\hspace{5mm} {\cal L}=\int dr\ \frac{1}{r}\tilde{\phi}^{\da}(O\tilde{\phi})=\int du\ \tilde{\phi}^{\da}(u,x)(O\tilde{\phi}(u,x))
\end{equation}
Now we can apply the usual Kaluza-Klein compactification procedure:
\begin{equation}
u\approx u+2\pi
\end{equation}
Since at least for the bosons we have the flat space-time action in terms of real gauge fields, it is perhaps useful to expand the gauge fields in terms of sines and cosines (one can of course also use exponential modes):
\begin{equation}
\tilde{\phi}(u,x)=\sum_{m}[\phi_{1m}(x)cos(mu)+\phi_{2m}(x)sin(mu)]
\end{equation}
If one substitutes (59) in (57) one finds that after the $u$ integration the different ``$m$-modes'' separate, which is essentially guaranteed by scale invariance of the original action (54).  The decoupling of the sine and the cosine modes is  trickier and we will discuss that in the next section, but for now we turn to a more important matter, that of unitarity. 

For dS the above prescription goes through without any complications and after compactification we obtain a gauge invariant dS action for each separate mode which differ from each other in their ``mass parameter'' $m$. However, as we anticipated before, compactification of $u$ which is time-like in AdS leads to problems with unitarity  although we do obtain gauge invariant actions\footnote{Such actions have appeared in the literature.}. To solve this problem we invoke Wick rotation (before performing the compactification) so that $u$ becomes imaginary:
\begin{equation}
u'\equiv -iu;\ \ u'\approx u'+2\pi
\end{equation}
The Wick rotation has two implications in the subsequent compactification procedure. First, since the gauge fields are now periodic in $u'$, in terms of $u$ the mode expansion becomes
\begin{equation}
\tilde{\phi}(u,x)=\sum_{m}[\phi_{1m}(x)cosh(mu)+\phi_{2m}(x)sinh(mu)]
\end{equation}
This explains for example, why we needed to use the modes $e^{mu}$ instead of $e^{imu}$ for AdS to derive the field equations (41)-(43). Second, since $r=e^{u}=e^{iu'}$, integration of $u'$  from 0 to $2\pi$ is a contour integral in  $r$ around 0 in the complex r-plane.  It is then easy to check that for both dS and AdS we have (writing everything in terms of $r$ to avoid confusion)
\begin{equation}
\tilde{\phi}(r,x)=\sum_{m}[\phi_{1m}(x)C_m(r)+\phi_{2m}(x)S_m(r)]
\end{equation}
where $C_m(r)$ and $S_m(r)$ are the harmonic modes which satisfy the orthogonality condition
\begin{equation}
\int \frac{dr}{r} C_m(r)C_n(r)\sim \de_{mn};\hspace{5mm}  \int \frac{dr}{r} S_m(r)S_n(r)\sim \eta\de_{mn};\hspace{5mm}  \int \frac{dr}{r} C_m(r)S_n(r)=0;\  
\end{equation}
where for AdS the $r$-integral is a contour integral while for dS it is the usual line integral. We have omitted the precise normalizations because as noted before the overall coefficients in front of the reduced mode actions are unimportant. For later convenience we note 
\begin{equation}
\Delt C_m(r)=-m\eta S_m(r);\hspace{5mm} \Delt S_m(r)=m C_m(r)
\end{equation}
\vspace{5mm}
\\
{\bf Bosonic Actions:} We start with the flat space-time bosonic action that one obtains through $OSp(1,1|2)$ formalism\footnote{One could also begin with the action in IGL(1) formalism \cite{igl}.} \cite{flataction} using BRST techniques. In this method one first adds 4 extra coordinates, two bosonic (one space like and one time-like) and two fermionic, to the D+1 flat space-time coordinates. Then one performs a light-cone reduction which eliminates the two light-cone coordinates (from the D+3 bosonic dimensions) so that we are still left with a  D+1 dimensional bosonic space (which is now treated as the space-time manifold) and two fermionic dimensions. The gauge fields are a representation of this $(D+1|2)$ dimensional space-time. In this formalism the action reads after covariantization
\begin{equation}
\hat{S}=\int d\hat{x}\ \hat{e}\hat{{\cal L}_b};\hs \hat{{\cal L}}_b=\2\phi^{\da}\hat{K}\hph
\end{equation}
where
\begin{equation}
\hat{K}=\2(-\hn^2+\2\hat{Q}^{\al}\hat{Q}_{\al});\hs \hat{Q}^{\al}=S^{\al B}\hn_B
\end{equation}
and is invariant under the  gauge transformation: 
\begin{equation}
\de \hph=\2\de_{s0}\hat{Q}^{\al}\hat{\La}_{\al}
\end{equation}
where $\de_{s0}$ picks up the singlet piece with respect to $S_{\al\bb}$, where $\al,\bb$ are the fermionic indices. This is a first-order formalism where $\hph$, $\La_{\al}$ are treated as states in the Hilbert space which is a  representation for the spin operators $S_{\bar{A}\bar{B}},\ \bar{A}=(A,\al)$.

Now we already have the expression for $\hn^2$ (30) and  putting  $\om=0$ we have
\begin{equation}
\hn^2=\frac{1}{r^2}[\n^2-2\eta \n^aS_{ra}+\eta \la(D+\la-1) + S_{ra}S_r{}^{a}]
\end{equation}
where we have to replace $\la$
\begin{equation}
\la\ra-d+\Delt
\end{equation}
Substituting (69) in (68) we have
\begin{equation}
\hn^2\hph=r^{-d-2}[\n^2-2\eta \n^aS_{ra}+\eta(-d^2+\p^{\x 2}) + S_{ra}S_r{}^{a}]\tilde{\phi}
\end{equation}
After integration we have
$${\cal L}_1\equiv -\2\int dr\hph^{\da}\hn^2\hph$$
$$= -\2\sum_m\{\phi_{1m}^{\da}[\n^2-2\eta \n^aS_{ra}+\eta(-d^2-\eta m^2)+ S_{ra}S_r^{\ a})\phi_{1m}]$$
$$ +\eta\phi_{2m}^{\da}[\n^2-2\eta \n^aS_{ra}+\eta(-d^2-\eta m^2) + S_{ra}S_r^{\ a}]\phi_{2m}\}$$
\begin{equation}
\equiv \sum_m {\cal L}_{1m}
\end{equation}
Let us now evaluate $\hat{Q}^{\al}\hat{Q}_{\al}$.
$$\hat{Q}^{\al}\hat{Q}_{\al}=C_{\al\bb}S^{\al A}S^{\bb B}\hn_A\hn_B$$
$$=C_{\al\bb}(S^{\al a}S^{\bb b}\hn_a\hn_b+S^{\al a}S^{\bb r}\hn_a\hn_r+S^{\al r}S^{\bb b}\hn_r\hn_b+S^{\al r}S^{\bb r}\hn_r\hn_r)$$
We can evaluate the terms separately in the same way as we obtained $\hn^2$ while deriving field equations:
$$\hn_a\hn_b=\frac{1}{r^2}(\n_a\n_b+\eta_{(a}M_{rb)}+\eta\eta_{ab}\la+M_{rb}M_{ra})$$
$$\hn_a\hn_r=\hn_r\hn_a=\frac{1}{r^2}(\la-1)(\n_b+\eta M_{ra})$$
and $$\hn_r\hn_r=\frac{1}{r^2}\la(\la-1)$$
Thus we have (changing the second quantized operators to first)
$$\hat{Q}^{\al}\hat{Q}_{\al}=\frac{1}{r^2}[S^{\al a}S_{\al}{}^b\n_a\n_b-\eta S^{\al a}S_{\al}{}^bS_{r(b}\n_{a)}+(\la-1)[S^{\al r},S_{\al}{}^b]\n_b$$
\begin{equation}
+\eta\la S^{\al a}S_{\al a}+S^{\al a}S_{\al}{}^bS_{ra}S_{rb}-\eta(\la-1)[S^{\al r},S_{\al}{}^b]S_{rb}+\la(\la-1)S^{\al r}S_{\al}{}^r]
\end{equation}
Again substituting (19) in (72) we get 
$$\hat{Q}^{\al}\hat{Q}_{\al}\hph=r^{-d-2}[S^{\al a}S_{\al}{}^b\n_a\n_b-\eta S^{\al a}S_{\al}{}^bS_{r(b}\n_{a)}+(-d+\Delt-1)[S^{\al r},S_{\al}{}^b]\n_b$$
$$+\eta\la S^{\al a}S_{\al a}+S^{\al a}S_{\al}{}^bS_{ra}S_{rb}-\eta(-d+\Delt-1)[S^{\al r},S_{\al}{}^b]S_{rb}$$
\begin{equation}
+(-d+\Delt)(-d+\Delt-1)S^{\al r}S_{\al}{}^r]\tilde{\phi}
\end{equation}
After performing the integral we obtain
$${\cal L}_2\equiv \4\int dr\ \hph^{\da}\hat{Q}^{\al}\hat{Q}_{\al}\hph$$
$$=\4\sum_m\{\phi_{1m}^{\da}[S^{\al a}S_{\al}{}^b\n_a\n_b-\eta S^{\al a}S_{\al}{}^bS_{r(b}\n_{a)}+(-d-1)[S^{\al r},S_{\al}{}^b]\n_b$$
$$+\eta\la S^{\al a}S_{\al a}+S^{\al a}S_{\al}{}^bS_{ra}S_{rb}-\eta(-d-1)[S^{\al r},S_{\al}{}^b]S_{rb}$$
$$-(d(-d-1)+\eta m^2)S^{\al r}S_{\al}{}^r]\phi_{1m}$$
$$+\eta\phi_{2m}^{\da}[S^{\al a}S_{\al}{}^b\n_a\n_b-\eta S^{\al a}S_{\al}{}^bS_{r(b}\n_{a)}+(-d-1)[S^{\al r},S_{\al}{}^b]\n_b$$
$$+\eta\la S^{\al a}S_{\al a}+S^{\al a}S_{\al}{}^bS_{ra}S_{rb}-\eta(-d-1)[S^{\al r},S_{\al}{}^b]S_{rb}$$
$$-(d(-d-1)+\eta m^2)S^{\al r}S_{\al}{}^r]\phi_{2m}$$
$$+m\phi_{1m}^{\da}[[S^{\al r},S_{\al}{}^b]\n_b -\eta[S^{\al r},S_{\al}{}^b]S_{rb}+(-2d-1)S^{\al r}S_{\al}{}^r]\phi_{2m}$$
$$-m\phi_{2m}^{\da}[[S^{\al r},S_{\al}{}^b]\n_b -\eta[S^{\al r},S_{\al}{}^b]S_{rb}+(-2d-1)S^{\al r}S_{\al}{}^r]\phi_{1m}]\}$$
\begin{equation}
\equiv \sum_m {\cal L}_{2m}
\end{equation}
Thus the reduced action for a single mode is given by
\begin{equation}
S=\int dx\ e {\cal L}_b;\hs {\cal L}_b={\cal L}_{1m}+ {\cal L}_{2m}
\end{equation}
The original gauge transformations look like
$$\de \hph=\2\de_{s0}\hat{Q}^{\al}\hat{\La}_{\al}=\2\de_{s0} r^{-d}[S^{\al b}(\n_b+\eta M_{rb})+(-d+1+\Delt)S^{\al r}]\tilde{\La}_{\al}$$
where $\hat{\La}_{\al}\equiv r^{-d+1}\tilde{\La}_{\al}$. Comparing the coefficients of the harmonic modes we obtain
$$
\de \phi_{1m}=\2\de_{s0}[(S^{\al b}(\n_b-\eta S_{rb})+(-d +1)S^{\al r})\La_{1m\al}+mS^{\al r}\La_{2m\al}]
$$
and 
\begin{equation}
\de \phi_{2m}=\2\de_{s0}[(S^{\al b}(\n_b-\eta S_{rb})+(-d +1)S^{\al r})\La_{2m\al}-\eta mS^{\al r}\La_{1m\al}]
\end{equation}
As is clear from the form of the dS/AdS action and the gauge transformations, the decoupling of the ``sine'' and ``cosine'' mode is not obvious and we do not yet have a  proof that a diagonalization is possible in the most general case of arbitrary spin representations. However, since we know that there exists actions for dS/AdS in terms of a single real field it will be highly unnatural if the ``complex action'' that we have obtained does not contain it.
We believe that an underlying symmetry principle will make this decoupling possible, just as $CP$ (product of complex conjugation and parity)  enables this decoupling in the ordinary dimensional reduction. 

One can in general define the reality condition on fields by
$$\phi^{\da}=A\phi$$
where $A$ is usually an anti-unitary operator. In ordinary field theory $A$ is just the complex conjugation operator $C$. However, in the ordinary linear dimensional reduction we have to use $A=CP$, where $P$ is the parity operator, to obtain actions in terms of real fields. For Wick rotated field theory a similar reality condition, known as Osterwalder-Schrader reality condition, is used \cite{t}, where $A=T$. The analogue of parity for radial reduction is inversion ($u\ra -u \Leftrightarrow r\ra \frac{1}{r})$ which was successfully used (i.e. $A=CU$, $U$ being the inversion operator) to define reality conditions for conformal fields \cite{inversion}. Unfortunately,  we cannot use inversion since generically we are dealing with non-conformal theories which are not invariant under it. However, in the simple examples that we will study in the next section we will see the decoupling explicitly.
\vspace{5mm}
\\
{\bf Fermionic Actions:} Reduction of the fermionic action is slightly trickier than its bosonic counterparts essentially because of funny hermiticity properties of the gamma matrices. Like the bosonic counterparts the fermionic fields $(\hph_I)$ carry a representation of the spin operators ($S_{\bar{A}\bar{B}}$). Moreover their index structure can be decomposed $I\ra (i,\al)$ where the  spinorial indices ($\al$) also carry a representation of the gamma matrices\footnote{Generically it is also possible to decompose  $S_{\bar{A}\bar{B}}$ into a vector and a spinorial part $S_{\bar{A}\bar{B}}=S_{v\bar{A}\bar{B}}+S_{s\bar{A}\bar{B}}$ where the vector part acts only on the $i$ index while the spinorial part acts on the $\al$ index.} \cite{flatfermions}. To start we have the D+1 dimensional flat space-time action given by
\begin{equation}
\hat{S}_f=\int d\hat{x}\ \hat{e}\hat{{\cal L}}_f;\hs \hat{{\cal L}}_f=\2\bar{\hph}\hat{K}_f\hph
\end{equation}
where
\begin{equation}
\hat{K}_f=\2\ga^{\al}S_{\al}{}^A\hn_A
\end{equation}
and $\bar{\phi}$ contains a product of gamma matrices in the time-like directions:
\begin{equation}
\bar{\phi}=\phi^{\da}\Upsilon;\hs  \Upsilon=\prod_{\eta_{AA}=-1}\sqrt{2}\ga^A
\end{equation}
Thus
\begin{equation}
\Upsilon_{dS}=\sqrt{2}\ga^0;\hs \Upsilon_{AdS}=2\ga^0\ga^r
\end{equation}
The gamma matrices satisfy the  anti-commutation relations
\cite{flatfermions}
\begin{equation}
\{\ga_A,\ga_B\}=(-1)^{t}\eta_{AB}
\end{equation}
where $t$ is the number of time dimensions in the manifold. Since we add an extra time dimension for AdS but an extra space dimension for dS, the anti-commutation relation looks like 
$$
\{\ga_A,\ga_B\}=-\eta\eta_{AB}
$$

The gauge transformations are given by (67), same as for the bosons. 

It is relatively easy to obtain the dimensionally reduced action applying the usual techniques:
$$S_f=\int dx\ e{\cal L}_f;$$
$$\ {\cal L}_f=\4[\bar{\phi}_{1m}\ga^{\al}(S_{\al}{}^{b}\n_b-\eta S_{\al}{}^{b}S_{rb}-d S_{\al}{}^{r})\phi_{1m}+\eta\bar{\phi}_{2m}\ga^{\al}(S_{\al}{}^{b}\n_b-\eta S_{\al}{}^{b}S_{rb}-d S_{\al}{}^{r})\phi_{2m}
$$
\begin{equation}
+m\bar{\phi}_{1m}\ga^{\al}S_{\al}{}^{r}\phi_{2m}-m\bar{\phi}_{2m}\ga^{\al}S_{\al}{}^{r}\phi_{1m}]
\end{equation}
Note that $\bar{\phi}$ still contains an extra factor of $\sqrt{2}\ga^r$ for AdS. The gauge transformation looks exactly the same as for the bosons (76). 

\section{SIMPLE EXAMPLES}

We have in the earlier section obtained both bosonic and fermionic actions in dS/AdS space-time in terms of gauge fields which are representations of $SO(D,1|2)$ or $SO(D-1,2|2)$, and  in case of fermions, also that of gamma matrices. The action (71),(74),(75) and (82) contains the spin operators (matrices) and one can work out the action for the special cases substituting the explicit spin matrices. However, often the flat space time action is known explicitly (i.e. there are no longer any spin matrices left) for a particular spin representation and in practice it is often easier to just dimensionally reduce this flat space-time action to obtain the dS/AdS action. For the simple examples that we choose we will adopt this strategy. 
\vspace{5mm}
\\
{\bf S=0, The Scalar Multiplet:} The flat space-time action is given by
\begin{equation}
\hat{S}_0=-\4\int d\hat{x}\ \hat{e}(\hn^A\hph)(\hn_A\hph)
\end{equation}
Using (19) we obtain
\begin{equation}
\hat{S}_0\equiv -\4\int dx\ e {\cal L}_0;\hs {\cal L}_0=\int\frac{dr}{r}\{(\n^a\tilde{\phi})^2+\eta[(-d+\Delt)\tilde{\phi}]^2)\}
\end{equation}
Extracting a single mode we find the dS/AdS action for a real field $\phi_{jm}$:
\begin{equation}
S_{0}\equiv -\4\int dx\ e {\cal L}_{0m};\hs {\cal L}_{0m}=(\n^a\phi_{jm})^2+ M_0^2(\phi_{jm})^2
\end{equation}
where $j$ runs over 1 and 2\footnote{There is a relative $-$ sign between the actions for $j=1$ and $j=2$ for AdS} and we have defined
\begin{equation}
M_0^2=m^2+\eta d^2
\end{equation}

Dimensional reduction of the scalar (multiplet) proved  trivial which was expected; since scalars do not possess any gauge invariance just a covariantization of the flat action should give us the dS/AdS action. Notice the cosmological contribution to the mass term differs by a sign for dS and AdS, a well-known fact.
\vspace{5mm}
\\
{\bf S=1, The Vector Multiplet:} The flat space-time action is given by
\begin{equation}
\hat{S}_1=-\4\int d\hat{x}\ \hat{e}\hat{F}^2;\hs \hat{F}_{AB}=\hn_{[A}\hph_{B]}
\end{equation}
Using (19) we can obtain the action in terms of ``Weyl transformed'' fields:
\begin{equation}
\hat{F}_{ab}=r^{-d-1}\tilde{F}_{ab};\hs \tilde{F}_{ab}\equiv\n_{[a}\tilde{\phi}_{b]}
\end{equation}
\begin{equation}
\hat{F}_{ar}\equiv r^{-d-1}\tilde{F}_{ar}=r^{-d-1}[\n_a\tilde{\phi}-(-d+1+\Delt)\tilde{\phi}_a]
\end{equation}
and
\begin{equation}
\hat{F}^2=\hat{F}_{ab}\hat{F}^{ab}+2\hat{F}_{ar}\hat{F}^{ar}
\end{equation}
We can concentrate on a single $m$ value as they decouple in an obvious manner. We have, dropping the subscript $m$,
$$\hat{F}_{ar}=r^{-d-1}\{C[\n_a\phi_{1}-((-d+1)A_{1a}+mA_{2a})] + S[\n_a\phi_{2}-((-d+1)A_{2a}-\eta mA_{1a})]\}$$
where we have defined 
\begin{equation}
\phi_{a}\equiv A_{a};\hs \phi_{r}\equiv \phi
\end{equation}
Let us further  define new fields $A'$ by
$$A'_{a}\equiv U_1 A_a\equiv \frac{1}{M_1}\left( \begin{array}{cc} -d+1&m\\
                                             -\eta m &-d+1
\end{array} \right)A_a
;\hs A_a\equiv \left( \begin{array}{c} A_{1a}\\
                                       A_{2a}
\end{array} \right);$$
\begin{equation}
M_1^2=\eta(-d+1)^2+m^2
\end{equation}
Then
$$\hat{F}_{ar}=r^{-d-1}[C(\n_a\phi_1-M_1A'_{1a}) + S(\n_a\phi_2-M_1A'_{2a})]$$
We can now perform the $r$ integral to obtain the mode action:
$$\int \frac{dr}{r}(\tilde{F}_{ab}\tilde{F}^{ab}+2\tilde{F}_{ar}\tilde{F}^{ar})$$
$$\sim [(F^2_{1}+\eta F^2_2)+2\eta((\n_a\phi_1-M_1A'_{1a})^2+\eta(\n_a\phi_2-M_1A'_{2a})^2)]$$
$$=\eta(F'^2_{1}+\eta F'^2_2)+2\eta[(\n_a\phi_1-M_1A'_{1a})^2+\eta(\n_a\phi_2-M_1A'_{2a})^2]$$
$$=2\eta\{[(\n_aA_{1b})^2-(\n\cdot A_1)^2+\eta(\n_aA_{2b})^2-(\n\cdot A_2)^2]+\eta d(A_1^{'2}+\eta A_2^{'2})$$
$$+[(\n_a\phi_1-M_1A'_{1a})^2+\eta(\n_a\phi_2-M_1A'_{2a})^2]\}$$
where we have used 
\begin{equation}
[\n_a,\n_b]=\eta M_{ab}
\end{equation}
Thus finally we obtain the dS/AdS action as 
$$S_{1}\equiv -\2\int dx\ e {\cal L}_{1};$$
\begin{equation}
{\cal L}_{1}=(\n_aA'_{jb})^2-(\n\cdot A'_j)^2+(\n_a\phi'_j)^2+2M_1\phi'_j\n\cdot A'+(M_1^2+\eta d)A_j^{'2}
\end{equation}
We now look at the gauge invariance. The original gauge transformations are given by
\begin{equation}
\de\hph_A=\hn_A \hat{\La}
\end{equation}
which translates to 
$$\de \tilde{\phi}_a=\n \tilde{\La};\hs \de \tilde{\phi}_r=(-d+1+\Delt)\tilde{\La}$$
Harmonic expansion then give us the dS/AdS gauge transformations which in matrix notation looks like:
\begin{equation}
\de A_a=\n \La;\hs \de\phi=M_1U_1\La
\end{equation}
or 
\begin{equation}
\de A'_a=\n \La';\hs \de\phi=M_1\La';\hs \La'\equiv U_1\La
\end{equation}
Our results coincide with the known results for the vector multiplet on dS/AdS. The vector multiplet thus generically consists of a massive vector and a massless scalar. However, observe that for the special value of $M_1=0$ the vector and the scalar dissociate and thus even though the ``mass term''  of the vector is non-zero, this special case is referred to as the massless vector multiplet. It should perhaps be mentioned that for the vector multiplet we have only one such ``critical value'' for the mass parameter. However, for higher multiplets there is also a phenomenon of ``partial masslessness''  which occurs at other critical values of the mass parameter, as we shall see in the next example of spin 2. ``Partial massless'' multiplets were first discovered in \cite{deser} after which a lot of progress have been made towards understanding the phenomena of partial masslessness and unitarity in (A)dS multiplets. For details, please see \cite{adsmass} and references therein.  Note that we can see the decouplings, both between the ``sine'' and the ``cosine'' modes in the generic case and between the scalar and the vector in the massless limit, at the level of gauge transformations themselves. This is convenient because gauge transformations are easy to study and, for example, issues like whether it is always possible to decouple the highest spin field from the rest of the multiplet even in the massless limit for any arbitrary representation \cite{adsmass} can be addressed.   
\vspace{5mm}
\\
{\bf S=2, The Graviton:} For the graviton we satisfy ourselves by looking only at the gauge transformations. As we will discover it provides us with a significant insight into the multiplet. The flat D+1 dimensional action is given by
\begin{equation}
\hat{S}_2=-\4\int d\hat{x}\ \hat{e}[\hph^{AB}\hn^2\hph_{AB}+2(\hn^B\hph_{AB})^2-\hph\hn^2\hph+2\hph\hn^B\hn^C\hph_{BC}]
\end{equation}
which is invariant under the gauge transformations
\begin{equation}
\de \hph_{AB}=\hn_{(A}\hat{\La}_{B)}
\end{equation}
In terms of the ``Weyl transformed'' fields we have
$$\de \tilde{h}_{ab}=\n_{(a}\tilde{\La}_{b)}+2\eta\eta_{ab}\tilde{\La}$$
$$\de \tilde{A}_a=\n_a\tilde{\La}+(-d+\Delt)\tilde{\La}_a$$
\begin{equation}
\de\tilde{\phi}=2(-d+1+\Delt)\tilde{\La}
\end{equation}
where we have defined
$$\hph_{ab}=r^{-d}\tilde{h}_{ab};\hs \hph_{ar}=r^{-d}\tilde{A}_a;\hs \hph_{rr}=r^{-d}\tilde\phi;$$
\begin{equation}
\hat{\La}_a=r^{-d+1}\tilde{\La}_a;\hs \hat{\La}_r=r^{-d+1}\tilde{\La} 
\end{equation}
As usual harmonic expansions give us the gauge transformations for the harmonic modes, which in matrix notation look like
$$\de h_{ab}=\n_{(a}\La_{b)}+2\eta\eta_{ab}\La;\hs \de A_a=\n_a\La+M_2U_2\La_a;\hs
\de\phi=2M_1U_1\La$$
where
\begin{equation}
M_2^2=\eta d^2+m^2;\hs M_1^2=\eta (-d+1)^2+m^2
\end{equation}
and 
\begin{equation}
U_1 \equiv \frac{1}{M_1}\left( \begin{array}{cc} -d+1&m\\
                                             -\eta m &-d+1
\end{array} \right)
;\hs U_2 \equiv \frac{1}{M_2}\left( \begin{array}{cc} -d&m\\
                                             -\eta m &-d
\end{array} \right)
\end{equation}
It is now easy to diagonalize the system. We define
\begin{equation}
h'_{ab}\equiv h_{ab}-\frac{U_1^{-1}}{M_1}\eta\eta_{ab}\phi;\hs A_a=U_2A'_a;\hs \phi\equiv U_1U_2\phi';\hs \La\equiv U_2\La'
\end{equation}
In terms of the prime fields the system decouples:
\begin{equation}
\de h'_{ab}=\n_{(a}\La_{b)};\hs \de A'_a=\n_a\La'+M_2\La_a;\hs
\de\phi'=2M_1\La'
\end{equation}
with
\begin{equation}
M_1^2=M_2^2-\eta[d^2-(-d+1)^2]=M_2^2-\eta(2d-1)
\end{equation}
This is a familiar form for the gauge transformations of the spin two multiplet in dS/AdS background (see for example \cite{adsbosact}). We can immediately make several observations. First, we observe that the sine and the cosine modes decouple as we conjectured. Second, we can obtain the critical values for the masslessness/partial masslessness conditions. From the gauge transformations (102) it is clear that at $M_2=0$ the multiplet dissociates into two sets, the graviton with gauge parameter $\La_a$ and the vector multiplet with gauge parameter $\La$. This obviously corresponds to the massless limit. We also find a second critical point for dS at $M_1=0$ and $M_2^2=(2d-1)$ when the scalar decouples from the graviton and the vector \cite{deser}. The residual graviton multiplet is invariant under both $\La_a$ and $\La$ gauge transformations. From (102) it is also clear that for certain range of values the theory will become non-unitary \cite{adsmass,adsbosact}. For example, in dS if $M_2^2<(2d-1)=D-2$, then unitarity is violated. 

To summarise we have shown that the sine and cosine modes decouple in the graviton multiplet, so that using our prescription one can obtain the AdS/dS action for the graviton multiplet. Further just analysing the gauge transformations provides us with insight with regards to issues like unitarity, masslessness/partial masslessness etc.  
\vspace{5mm}
\\
{\bf S = 1/2, The Matter Multiplet:} As we mentioned before, the issue of fermions is a little tricky specifically in AdS. This is because in $\bar{\hph}$ we have an extra gamma matrix corresponding to the extra time-like direction but we obtain the correct action once we realize that the extra gamma matrix $\ga_r$ plays the role of ``$\ga_5$'' or the chirality operator. We start with the flat space-time action given by
\begin{equation}
\hat{S}_{\2}=\int d\hat{x}\ \hat{e}\hat{{\cal L}}_{\2};\hs \hat{{\cal L}}_{\2}=i\4\bar{\hph}\ga^A\hn_A\hph 
\end{equation}
Now
$$\hat{{\cal L}}_{\2}=i\4\bar{\hph}(\ga^a\hn_a\hph +\ga^r\hn_r\hph)= i\frac{1}{4}r^{-2d-1}\bar{\tilde{\phi}}[\ga^a\n_a -\eta\ga^a S_{ra}+(-d+\Delt)\ga^r]\tilde{\phi}$$
A standard way to deal with fermions \cite{flatfermions} is to dissociate the spin operator into the tensor and the spinorial part:
\begin{equation}
S_{AB}=S_{AB}^v+S_{AB}^s;\ S_{AB}^s=\2(-1)^t[\ga_A,\ga_B]
\end{equation}
where $(-1)^t$ is also $-\eta$. For spin half the tensor spin vanishes and thus we have after some playing around with the gamma matrices
$$\hat{{\cal L}}_{\2}=i\4\bar{\tilde{\phi}}(\ga^a\n_a+\eta\ga_r\Delt)\tilde{\phi}$$
From this point on we have to deal with the AdS and dS separately.

For dS, the gamma matrices satisfy
\begin{equation}
\{\ga_a,\ga_b\}=-\eta_{ab};\hs \ga^{\ad\da}=-\ga^{\ad};\hs \ga^{0\da}=\ga^0;\hs \ga^{r\da}=-\ga^r
\end{equation}
where $\ad=1...D-1$.
Now expanding the action in the harmonic modes and keeping only the terms for a single $m$ we have
\begin{equation}
{\cal L}_{\2}\equiv\int dr\ r^{D}\hat{\cal L}_{\2}=i\4[\bar{\phi}_1\ga^a\n_a\phi_1+\bar{\phi}_2\ga^a\n_a\phi_2+m(\bar{\phi}_1\ga_r\phi_2-\bar{\phi}_2\ga_r\phi_1)]
\end{equation}
$\ga_r$ is an anti-hermitian operator (at least for even D we can identify $i\ga_r$ as the chirality operator) and we can choose
\begin{equation}
\ga_r\phi_1=\textstyle{i\over \sqrt{2}}\phi_1 \ \mbox{ and }\  \ga_r\phi_2=-\textstyle{i\over \sqrt{2}}\phi_2
\end{equation}
We could also have chosen the opposite configuration, these two sets clearly decouple in the action. With (111) we obtain
\begin{equation}
 {\cal L}_{\2}=i\4[\bar{\psi}_L\ga^a\n_a\psi_L+\bar{\psi}_R\ga^a\n_a\psi_R-im\textstyle{1\over \sqrt{2}}(\bar{\psi}_L\psi_R+\bar{\psi}_R\psi_L)]
\end{equation}
where we have introduced\footnote{The subscripts ``$L$'' and ``$R$'' does suggest its connection to chirality but this connection is in no way neccessary and in particular for odd D they just denote the positive and negative eigenvalues of $\ga_r$.}
\begin{equation}
\psi_L\equiv\phi_1 \ \mbox{ and }\ \psi_R\equiv\phi_2
\end{equation}
One immediately recognises the Dirac Lagrangian:
$${\cal L}_{\2}=i\4[\bar{\psi}_D\ga^a\n_a\psi_D-m\textstyle{i\over \sqrt{2}}\bar{\psi}_D\psi_D]$$
where $\psi_D=\psi_L+\psi_R$. There is however a small subtelty when D is odd. Since the Dirac spinor in D+1 dimensions has twice the components as the Dirac spinor in D dimensions (for odd D) the Dirac Lagrangian that we obtain through dimensional reduction is reducible (the same is true for ordinary dimensional reduction). However, it is easy to see that this reducible spinor decomposes into two D-dimensional Dirac spinors. For example, one can start with a representation of the D+1 dimensional gamma matrices such that the $\ga^a$'s are block diagonal. Another strategy to deal with odd dimensions could be to start with Weyl spinors in D+1 dimensions, which has the same number of components as the D-dimensional Dirac spinor and would presumably reduce to it.

For AdS, the gamma matrices satisfy
\begin{equation}
\{\ga_a,\ga_b\}=\eta_{ab};\hs \ga^{\ad\da}=\ga^{\ad};\hs \ga^{0\da}=-\ga^0;\hs \ga^{r\da}=-\ga^r
\end{equation}
where as before $\ad=1...D-1$.
Now expanding the action in the harmonic modes and keeping only the terms for a single $m$ we have
\begin{equation}
{\cal L}_{\2}\equiv\int dr\ r^{D}\hat{{\cal L}}_{\2}=i\4[\bar{\phi}_1\ga^a\n_a\phi_1-\bar{\phi}_2\ga^a\n_a\phi_2+m(\bar{\phi}_1\ga_r\phi_2-\bar{\phi}_2\ga_r\phi_1)]
\end{equation}
We should remember however that the ``bar'' contains an extra factor of $\ga_r$ and hence converting them to normal ``bar'' we obtain
\begin{equation}
{\cal L}_{\2}= i\4[\sqrt{2}(\bar{\phi}_1\ga^r\ga^a\n_a\phi_1-\bar{\phi}_2\ga^r\ga^a\n_a\phi_2)+\textstyle{1\over \sqrt{2}} m(\bar{\phi}_1\phi_2-\bar{\phi}_2\phi_1)]
\end{equation}
Note that $\ga_r$ is still anti-hermitian (114) and we can choose
\begin{equation}
\ga_r\phi_1=\textstyle{i\over \sqrt{2}}\phi_1 \ \mbox{ and }\  \ga_r\phi_2=-\textstyle{i\over \sqrt{2}}\phi_2
\end{equation}
As in dS we could also have chosen the opposite configuration. We obtain
$${\cal L}_{\2}= i\4[i(\bar{\phi}_1\ga^a\n_a\phi_1+\bar{\phi}_2\ga^a\n_a\phi_2)+\textstyle{1\over \sqrt{2}} m(\bar{\phi}_1\phi_2-\bar{\phi}_2\phi_1)]$$
$$=i\4[i(\bar{\psi}_L\ga^a\n_a\psi_L+\bar{\psi}_R\ga^a\n_a\psi_R)-\textstyle{i\over \sqrt{2}}m(\bar{\psi}_L\psi_R+\bar{\psi}_R\psi_L)]$$
where we have introduced
\begin{equation}
\psi_L=\phi_1 \ \mbox{ and }\ \psi_R=i\phi_2
\end{equation}
Finally we should remember that the gamma matrices $\ga_a$ correspond to the D+1 dimensional flat space-time (with two time-like directions) and hence have the ``wrong'' sign in their anti-commutation relations and  satisfy the opposite hermiticity properties as compared to the conventions we prescribed (79), (81) and (108) for D dimensional AdS which obviously has only one time-like direction. If we define
\begin{equation}
\ga'_a=-i\ga_a
\end{equation}
we obtain the ``right'' gamma matrices and the Lagrangian is then given by
\begin{equation}
{\cal L}_{\2}\sim i\4[(\bar{\psi}_L\ga'^a\n_a\psi_L+\bar{\psi}_R\ga'^a\n_a\psi_R)+\textstyle{i\over \sqrt{2}}m(\bar{\psi}_L\psi_R+\bar{\psi}_R\psi_L)]
\end{equation}
Again we recognise the Dirac Lagrangian, aside from the subtelty for odd dimensions which can be taken care of in  exactly the same way as in dS.

\section{INTERACTIONS}

So far we have been discussing radial dimensional reduction of free theories. However, the success of the method is tempting for us to try and generalise this procedure to interacting theories.

 To relate the radial reduction to the usual Kaluza-Klein reduction it is convenient to work in terms of the Weyl transformed fields (19) and introduce the coordinate $u=ln (r)$ as we did for the flat case, with $u\ \e (-\infty,\infty)$ as $r\ \e [0,\infty)$. Kaluza-Klein reduction now amounts to making the $u$ coordinate periodic:
\begin{equation}
u\sim u+2\pi U
\end{equation}
The various fields can now be expressed as a harmonic expansion in the $u$ coordinate. For example for a general matter field we have
\begin{equation}
\tilde{\phi}(u,x)=\sum_{n} \phi_n(x)e^{inu/U}
\end{equation}
which in terms of the $r$ coordinate implies
\begin{equation}
\tilde{\phi}(r,x)=\sum_{n} \phi_n(x)r^{in/U}
\end{equation}
where
\begin{equation}
\hph(r,x)=r^{-d}\tilde{\phi}(r,x)
\end{equation}
The radial dimensional reduction that we discussed in the earlier section clearly corresponds to choosing a particular harmonic ($m=\frac{n}{U}$). It is clear that just as we required that the D+1 dimensional flat space-time free theories to be scale invariant (we only considered massless theories)  for the dimensional reduction to work,  we will now need the interacting theories to be scale invariant for the same reasons. Further, since we are now dealing with an interacting theory, we have to use a $U(1)$ symmetry generator \cite{u1}, say $G$, to select the eigenmodes corresponding to the different fields of the interacting theory, which will combine to form the D dimensional multiplet. In the conventional dimensional reduction scheme this is acheived by modifying the eigenvalue equation $P_D=m$ that was used for free theories to $P_D=mG$. For the radial reduction it works almost the same way, except that we have to replace $P$ by $\Delta$ just as we did for free threories (1).
\begin{equation}
\Delta = imG \ \hbox{\hspace{5mm} instead of \hspace{5mm}} \ P_D = mG
\end{equation}
This ensures that the ``mass phases'' coming from the eigenfunctions of the different fields cancel in each of the terms in the action, provided $G$ is a symmetry of the original action.

\section{SUMMARY AND FUTURE RESEARCH}

In this paper we have successfully obtained first the field equations (for field strengths) and then the action for the most general spin representations in dS/AdS of arbitrary space-time dimension. The success relied essentially on two key ingredients. First, the geometric picture of viewing dS/AdS as a ``sphere'' in one higher dimensional flat space-time of appropriate signature. This introduced the concept of ``radial dimensional reduction'', which differs from ordinary dimensional reduction by the fact that we end up with the ``sphere'' instead of a flat manifold. Second, was the ingredient of scale invariance which the flat space-time field equations and actions possessed. This helped us to get the algebra of the symmetry generators and field equations right and in case of actions guaranteed the decoupling of the different ``m modes''. This approach not only gives us the general results but also  provides us with a general algorithm to link between the flat  and dS/AdS space-time physics, which now hopefully will find many other applications. 

We discussed how the radial reduction of  ``free theories''  (the action being quadratic in the fields) can be generalized to include interactions. Another important generalization could be to consider radial dimensional reduction of an arbitrary manifold, essentially a supergravity theory. However, only future research will be able to tell for example, whether radial reduction of supergravity in 11 dimensions give a different (from the one that is obtained by ordinary dimensional reduction) consistent supergravity  in 10 dimensions. It is clear that for both these scenarios to work there has to be a $U(1)$ symmetry in the action which will be needed to preserve scale invariance (the imaginary mass parameters canceling against each other in every term in the action). This however may not be sufficient to guarantee the success of the reduction and of course one has to study these theories carefully (which we plan to do in future). One can think of other possible generalisations in the reduction scheme. For example, to apply similar techniques to obtain actions for non-maximally isometric spaces, or for that matter to reduce more than one dimension simultaneously. The problem with these are that we do not have any symmetry like scale invariance which was so crucial for the success of this reduction. Without further speculations it might just be fair to say that this algorithm that we have developed shows promise for future generalizations and applications. 

With regard to the specific problem we considered, even though we succeeded in our goal of  obtaining the most general action for dS/AdS some work still needs to be done to understand the dS/AdS multiplets better. Firstly, it will be nice to have a well-defined technique to obtain real multiplets from the complex multiplets for which we obtained the actions. Such a prescription exists for the usual reduction \cite{flataction}, so we hope an analogue will exist for radial reduction too. Secondly, and perhaps most importantly, we should try to understand the phenomenon of masslessness or partial masslessness in this framework.  A related question will be to predict which are the spin fields that one cannot dissociate from the highest spin-field in a spin multiplet even when mass is zero \cite{adsmass}. Thirdly, we know that the issue of unitarity is tricky in AdS and especially in dS. Previous research \cite{adsmass,adsbosact} has shown that there are regions in coupling parameters ($m$ in our case) where unitarity is violated. We do not yet know whether there is a geometric way to explain these regions. Finally it might be a good idea to obtain the various results that are already known for dS/AdS using this formalism which will also serve as a rigorous test for it. 
 
Thus as it stands there are a lot of open questions and possibilities that this work has provided us, hopefully future research will shed light upon them. 
  
\section{APPENDIX: Bosonic Field Equations in AdS}

We start with the field equations (33) and (34)\footnote{The $F_r$ equation is redundant because of linear dependence of the field equations.}. To compute them explicitly in the given reference frame we need to evaluate the covariant derivatives given by (13). As in section 2 we use Weyl transformations (14). Here we have
\begin{equation}
e_a{}^{m}=z\de_a{}^{m} \equiv W(z)e'_a{}^m
\end{equation}
Then ``prime'' covariant derivatives are just the partial derivatives, the ``prime'' vielbein being flat.   Thus the covariant derivatives are  given by
\begin{equation}
\n_{\ad}=z\p_{\ad}+M_{\ad z};\hs\n_z=z\p_z
\end{equation}
Using (127) one can compute the field equations:
\begin{equation}
F'=z^2\p^2-2zS_{\ad z}\p^{\ad}-(D-2)z\p_z+S^{\ad}_{\ z}S_{\ad z}-S_r^{\ a}S_{ra}+d^2-m^2
\end{equation}
\begin{equation}
F_a=zS_a^{\ b}\p_b-S_a^{\ \bd}S_{\bd z}+mS_{ra}+\2\{S_a^{\ b},S_{rb}\}
\end{equation}
It turns out that to compare with the known results we have to transform these field equation operators\footnote{One can view these as a rescaling of the spin-fields}:
\begin{equation}
F\ra z^{-n}Fz^n;\hs n=\frac{D-2}{2}
\end{equation}
After performing the rescaling the field equations look like
$$
\tilde{F}'=z^2\p^2-2zS_{\ad z}\p^{\ad}+S^{\ad}_{\ z}S_{\ad z}-S_r^{\ a}S_{ra}+\4-m^2
$$
$$
\tilde{F}_{\ad}=zS_{\ad}^{\ b}\p_b-S_{\ad}^{\ \bd}S_{\bd z}+mS_{r\ad}-nS_{\ad}^{\ z}+\2\{S_{\ad}^{\ b},S_{rb}\}$$ and
$$\tilde{F}_z=zS_z^{\ b}\p_b-S_z^{\ \bd}S_{\bd z}+mS_{rz}+\2\{S_z^{\ b},S_{rb}\}$$
After combining some equations and a little bit of algebra one can cast the equations into the following form:
\begin{equation}
\tilde{F}=\tilde{F}'-2\tilde{F}_z=z^2\p^2-A
\end{equation}
\begin{equation}
\tilde{F}_{\ad}=zS_{\ad}^{\ b}\p_b+C
\end{equation}
and
\begin{equation}
\tilde{F}_z=zS_z^{\ \bd}\p_{\bd}+B
\end{equation}
where
\begin{equation}
A=S^{\ad}_{\ z}S_{\ad z}+S^{\ad}_{\ r}S_{\ad r}+S^{z}_{\ r}S_{zr}+\{S^{\ \bd}_{z},S_{r\bd}\}+2mS_{rz}+m^2-\4
\end{equation}
\begin{equation}
B=-S^{\ \bd}_{z}S_{\bd z}+mS_{rz}+\2\{S^{\ \ad}_{z}S_{r\ad}\}
\end{equation}
and
\begin{equation}
C=\2[S_{\ad z},A]
\end{equation}
One can re-express $A$ and $B$ in terms of operators $(p_{\ad},k_{\ad},\de)$ as defined in (45) and after some tedious manipulations one gets (44).
\vspace{5mm}
\\
{\Large {\bf Acknowledgement}}
\vspace{5mm}

This work was supported in part by NSF Grant No. PHY-0098527.


\begin{thebibliography}{99}

\bibitem{adssusy} W. Nahm, {\it Nucl. Phys.} {\bf B135} (1978)
149
\bibitem{adsvacuum} M.A. Vasiliev, E.S. Fradkin, {\it Annal. Phys.} {\bf 177} (1987) 63; E.S. Fradkin, M.A. Vasiliev, {\it Phys. Lett.} {\bf B 189} (1987) 89; {\it Nucl. Phys.} {\bf B 291} (1987) 141; M.A. Vasiliev, {\it Annal. Phys.} {\bf 190} (1989) 59
\bibitem{adscft} J. Maldacena, {\it Adv. Theor. Math. Phys.} {\bf 2} (1998) 231; {\it Int. J. Theor. Phys.} {\bf 38} (1999) 1113,
\hhref{9711200}
\bibitem{dirac} P.A.M. Dirac, {\it Ann. Math.} {\bf 36} (1935) 657
\bibitem{ads4bosact} C. Fronsdal, {\it Phys. Rev.} {\bf D 18} (1978) 3624
\bibitem{vasbosact} V.Lopatin, M.Vasiliev, {\it Mod.Phys.Lett} {\bf A 3} (1988) 257
\bibitem{adsbosact} Y.M. Zinoviev, \hhref{0108192}
\bibitem{ads5bosact} R.R. Metsaev, \hhref{0201226} 
\bibitem{ads5feract} M.A.Vasiliev, {\it Nucl. Phys.} {\bf B 301} (1988) 26; K.B. Alkalaev, {\it Phys. Lett.} {\bf B 519} (2001) 121, 
\hhref{0107040}
\bibitem{flataction} W. Siegel, B. Zwiebach, {\it Nucl. Phys.} {\bf B 282} (1987) 125
\bibitem{flatmassive} O. Klein, {\it Z. Phys.} {\bf 37} (1926) 895; V. Fock, {\it Z. Phys.} {\bf 39} (1927) 226
\bibitem{deser} S. Deser, R.I. Nepomechie, {\it Phys. Lett.} {\bf B 132} (1983) 321; {\it Annals. Phys.} {\bf 154} (1984) 396
\bibitem{adsmass} L. Brink, R.R. Metsaev, M.A. Vasiliev,  {\it Nucl. Phys.} {\bf B 586} (2000) 183, \hhref{0005136}; S. Deser, A. Waldron, {\it Phys. Rev. Lett.} {\bf 87} (2001) 031601,\hhref{0102166}; {\it Nucl. Phys.} {\bf B 607} (2001) 577,
\hhref{0103198}
\bibitem{adslc} R.R. Metsaev, {\it Nucl. Phys.} {\bf B 563} (1999) 295,
\hhref{9906217}
\bibitem{flatbosact}  C. Fronsdal, {\it Phys. Rev.} {\bf D 18} (1978) 3624
\bibitem{flatferact}  C. Fronsdal, {\it Phys. Rev.} {\bf D 18} (1978) 3630
\bibitem{adsboseqn} M.A. Vasiliev, {\it Phys. Lett.} {\bf B 243} (1990) 378; {\it Phys. Lett.} {\bf B 257} (1991) 111; R.R. Metsaev, {\it Phys. Lett.} {\bf B 354} (1995) 78; 
\bibitem{adsfereqn} C. Fronsdal, {\it Phys. Rev.} {\bf D 18} (1978) 3624;
R.R. Metsaev, {\it Class. Quant. Grav.} {\bf 14} (1997) L115, \hhref{9707066}; {\it Phys. Lett.} {\bf B 419} (1998) 49, \hhref{9802097}
\bibitem{igl} W. Siegel, {\it Nucl. Phys.} {\bf B 288} (1987) 332
\bibitem{t} K. Osterwalder, R. Schrader, {\it Commun. Math. Phys.} {\bf 31} (1973) 83; {\bf 42} (1975) 281
\bibitem{inversion} S. Fubini, A.J. Hanson, R. Jackiw, {\it Phys. Rev.} {\bf D 7} (1973) 1732
\bibitem{flatfermions} W. Siegel, {\it Nucl. Phys.} {\bf 284} (1987) 632
\bibitem{u1} J. Scherk, J.H. Schwarz, {\it Phys. Lett.} {\bf B 82} (1979) 60
\end{thebibliography}
\end{document}